\documentclass[10pt,letterpaper]{article}
\usepackage{opex3}
\usepackage{float}
\usepackage{subfigure}
\usepackage{cite}

\newcommand{\beq}{\begin{equation}}
\newcommand{\eeq}{\end{equation}}
\newcommand{\barr}{\begin{array}}
\newcommand{\earr}{\end{array}}
\newcommand{\bea}{\begin{eqnarray}}
\newcommand{\eea}{\end{eqnarray}}

\newcommand{\mtwo}[3]{\left( \begin{array}{#1} #2 \\ #3 \end{array} \right)}

\newcommand{\mfour}[5]{\left( \begin{array}{#1} #2 \\ #3 \\ #4 \\ #5 \end{array} \right)}

\newcommand{\pd}[2]{\frac{\partial #1}{\partial #2}}

\newcommand{\lam}{\Lambda}
\newcommand{\half}{\frac{1}{2}}

\newcommand{\eps}{\epsilon}

\begin{document}

%%%%%%%%%%%%%%%%%% title page information %%%%%%%%%%%%%%%%%%
\title{Finite-Difference Time-Domain simulation of spacetime cloak}

\author{Jason Cornelius,$^{1}$  Jinjie Liu,$^{1,*}$ and Moysey Brio$^{2}$}

\address{1. Department of Mathematical Sciences, Delaware State University, Dover, DE 19901 \\
2. Department of Mathematics, The University of Arizona, Tucson, AZ 85721}

\email{$^*$ jliu@desu.edu} %% email address is required



\begin{abstract}
In this work, we present a numerical method that remedies the instabilities of the conventional FDTD approach  for solving Maxwell's equations in a space-time dependent magneto-electric medium with direct application to the simulation of the recently proposed spacetime cloak.
We utilize a dual grid FDTD method overlapped to the time domain to provide a stable approach for the simulation of magneto-electric medium with time and space varying permittivity, permeability and coupling coefficient.
The developed method can be applied to explore other new physical possibilities offered by spacetime cloaking, metamaterials, and transformation optics.
\end{abstract}

 % REPLACE WITH CORRECT OCIS CODES FOR YOUR ARTICLE
% 230.3205: Invisibility cloaks;
% 000.4430: Numerical approximation and analysis
% 160.3918: Metamaterials

%%%%%%%%%%%%%%%%%%%%%%% References %%%%%%%%%%%%%%%%%%%%%%%%%

%%%%%%%%%%%%%%%%%%%%%%%%%%  body  %%%%%%%%%%%%%%%%%%%%%%%%%%
\section{Introduction}

In recent years, Transformation Optics (TO) has become one of more interesting topics in science.
Utilizing transformation optics and metamaterials, spatial invisibility cloaks \cite{Pendry06,Leonhardt06} have been developed.  A spatial cloak functions by manipulating the spatial path of electromagnetic waves within the cloaked region in such a way that waves are bent around the object being cloaked.  Transformation optics allows for a description of the medium (ie: material parameters) needed for the construction of such devices.  For this particular type of cloak, the implementation requires the use of inhomogeneous media where the inhomogeneity is due to the use of spatially dependent anisotropic permeability and permittivity.

Within the past three years, a new type of cloak has been developed \cite{McCall11,Kinsler14}, the spacetime cloak.  Contrary to the spatial cloak, a spacetime cloak functions by altering the speed of the electromagnetic wave as it propagates through the cloaked region.  Initially, the speed of the front of the wave is increased while the speed of the back of the wave is decreased.  This allows for a gap to occur in which an event can be carried out undetected.  To close the gap, the back of the wave is sped up and the front is slowed down allowing the wave to continue on at its original speed.  The design of such a cloak can be achieved by transformation optics, and it results in a magneto-electric medium with time and space varying permittivity, permeability, and coupling coefficient.

Due to the interesting nature of these devices, simulation is desirable.  To this end, we turn to the Finite-Difference Time-Domain (FDTD) method.  The FDTD method \cite{Yee66,Taflove75,Taflove05} is a very successful method for simulating electromagnetic wave propagation.
It has been applied to the study of various types of materials \cite{Taflove05,Teixeira08}, including dielectrics, linear dispersive materials, nonlinear Raman and Kerr materials, nonlinear dispersive materials \cite{Liu10},
bi-isotropic media \cite{Akyurtlu04,Semichaevsky06}, etc.
In this paper, we are interested in space and time dependent magneto-electric materials and their direct application to the spacetime cloak.
In \cite{Kinsler14}, the FDTD method has been applied to simulate the spacetime cloak but the authors point out that there are some difficulties near the closing process of the spacetime cloak. Our numerical simulations demonstrate that there are instabilities for conventional FDTD simulation due to the temporal extrapolation of the time dependent magneto-electric constitutive equations.

To stably solve Maxwell's equations in time dependent magneto-electric medium,
we propose a modified FDTD method based on the use of time overlapped grids.
Our method uses two Yee grids that are offset in time by a half time step
so that collocated fields are provided in the time dependent magneto-electric constitutive equations. The resulting numerical update equations do not require time extrapolation, therefore the instability due to time extrapolation is avoided.  Our method is applied to simulate the spacetime cloak.

\section{Transformation Optics and spacetime cloak}
In this section, we briefly review the derivation of the spacetime cloak using transformation optics.
Consider the following covariant form of the Maxwell's equations (Amp\`ere's Law and the Gauss' Law of electric fields)
\beq
\tilde\nabla \tilde M=
\tilde \nabla
\mfour{cccc}{0&D_x&D_y&D_z}{-D_x&0&H_z&-H_y}{-D_y&-H_z&0&H_x}{-D_z&H_y&-H_x&0}
=0,
\label{eq:mxD}
\eeq
where $\tilde \nabla = (\pd{}{t},\pd{}{x},\pd{}{y},\pd{}{z})$.
Applying a spacetime transformation from $(t,x)$ to $(\tau,\xi)$ and keeping $y$ and $z$ coordinates unchanged, we have the following covariant form of the Maxwell's equation in the new coordinates $(\tau,\xi,\eta,\zeta)$:
\beq
\tilde\nabla' \tilde M' = \tilde\nabla' \mfour{cccc}{0&D_\xi&D_\eta&D_\zeta}{-D_\xi&0&H_\zeta&-H_\eta}{-D_\eta&-H_\zeta&0&H_\xi}{-D_\zeta&H_\eta&-H_\xi&0} = 0,
\eeq
where $\tilde\nabla'= (\pd{}{\tau},\pd{}{\xi},\pd{}{\eta},\pd{}{\zeta})$,
$\tilde M =|\tilde\Lambda|\tilde\Lambda^{-1} \tilde M' \tilde\Lambda^{-T}$,
and $\tilde \Lambda $ is the Jacobian matrix ($I_2$ represents the $2\times2$ identity matrix):
\beq
\tilde \Lambda = \frac{\partial(\tau,\xi,\eta,\zeta)}{\partial(t,x,y,z)} =
\mfour{cccc}{\tau_t&\tau_x&0&0}{\xi_t&\xi_x&0&0}{0&0&1&0}{0&0&0&1} =  \mtwo{cc}{\Lambda&0}{0&I_2}.
\eeq
Similar matrix equations are obtained for Faraday's Law together with the Gauss' Law of magnetic fields.

\begin{figure}[t]
 \centering
 \subfigure[]{\includegraphics[width=0.33\linewidth]{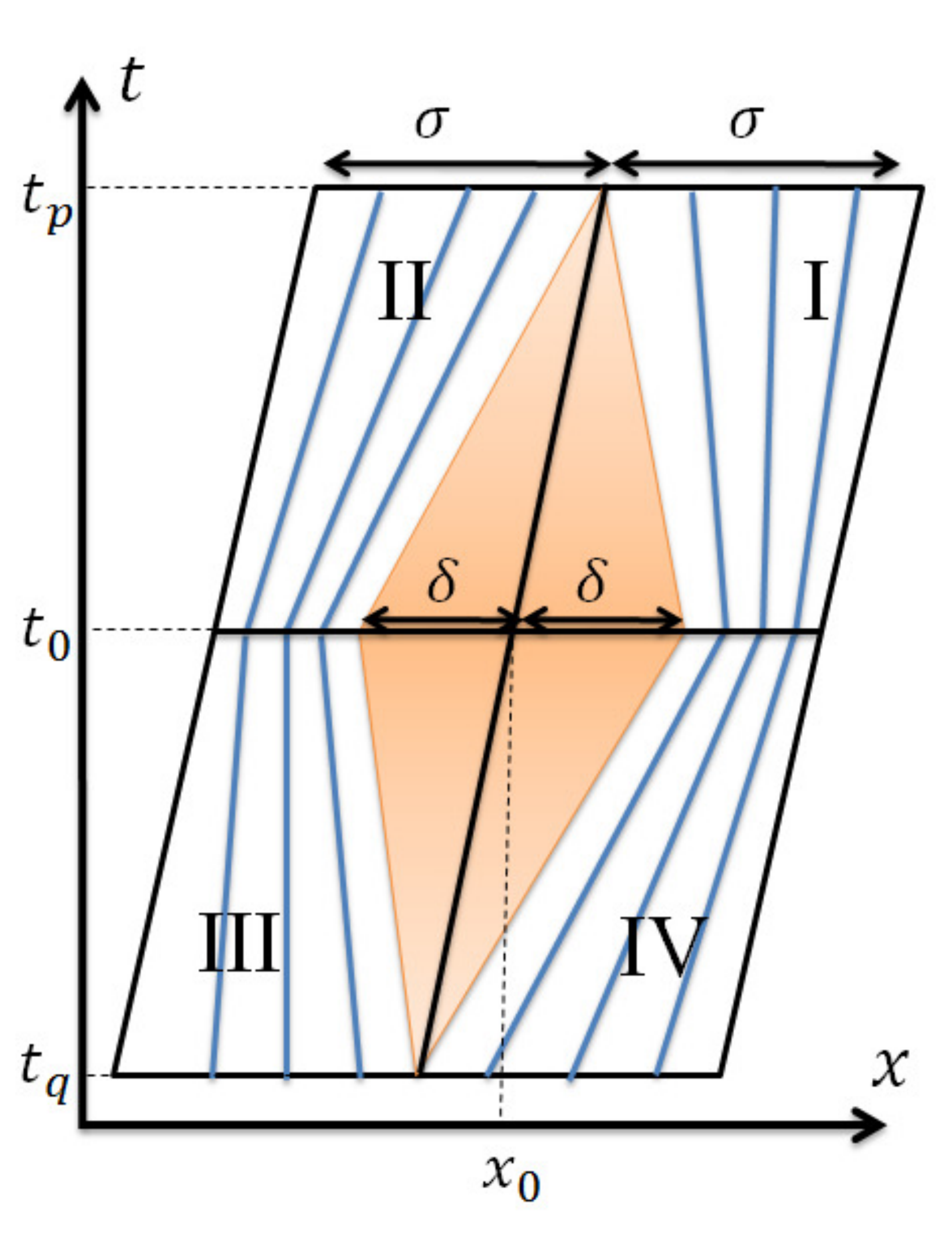} \label{fig:tranxt}} \hskip 20pt
 \subfigure[]{\includegraphics[width=0.33\linewidth]{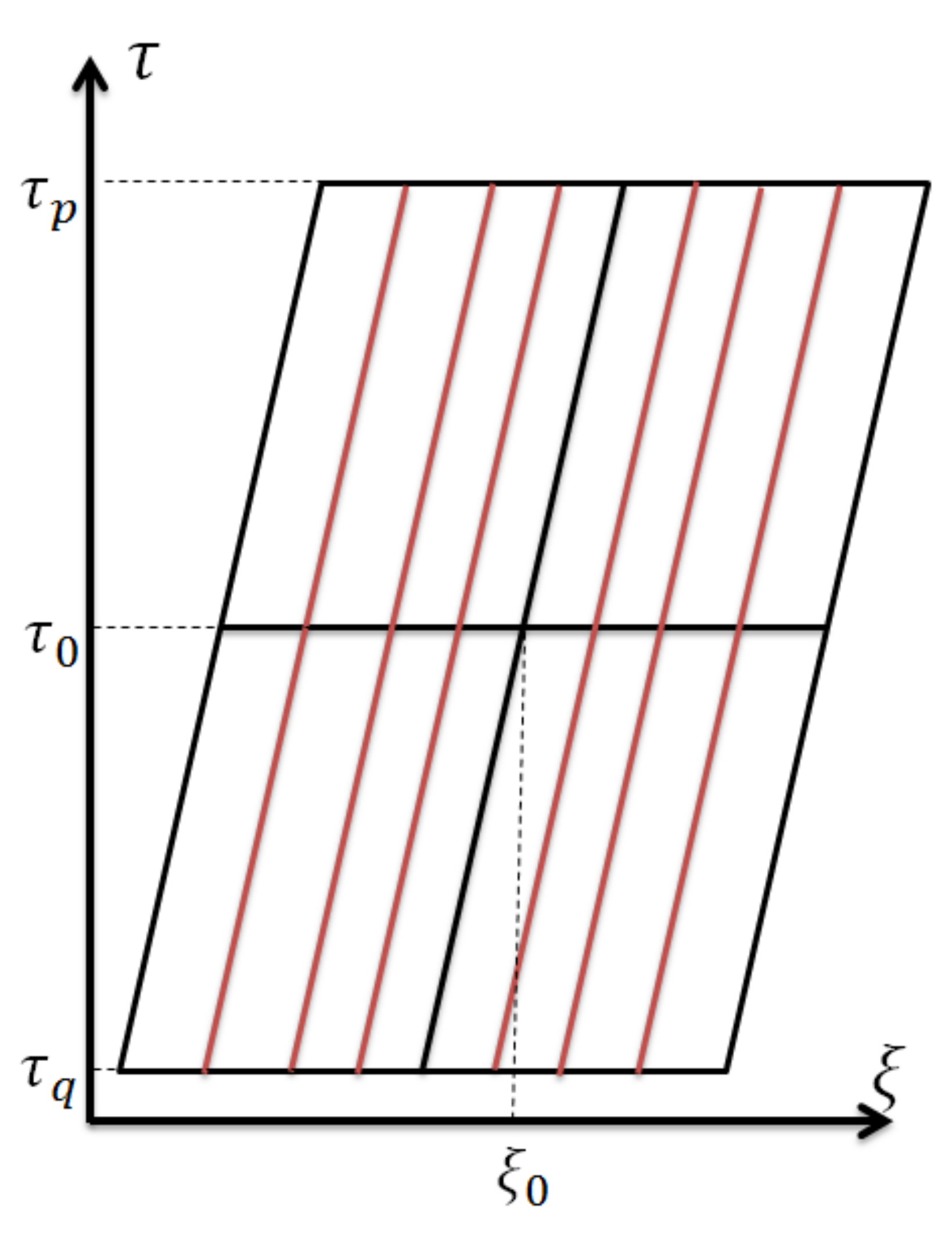} \label{fig:tranxitau}}
 \caption{(a) A spacetime cloak in the ($t,x$) domain. (b) Free space after the transformation to the ($\tau,\xi$) domain.}
\label{fig:tran}
\end{figure}

We consider a spacetime cloak design similar to the spacetime cloaks proposed in \cite{McCall11,Kinsler14}.  Figure~\ref{fig:tran} shows a diamond shaped spacetime cloak in the ($t, x$) domain and the transformed ($\tau, \xi$) free space. The original and the transformed regions are the same parallelograms centered at $(t_0, x_0)$ so we have $\xi_0 = x_0$, $\tau_0=t_0$, $\tau_p=t_p$, and $\tau_q=t_q$.
In the parallelogram we apply the following coordinate transformation to create such spacetime cloak:
\bea
\tau 	&=& t, \\
\xi 	&=& \frac{\sigma(x-x_S)}{\sigma - \delta(t-t_S)/(t_0-t_S)} + x_S,
\eea
where the values of $x_S$ and $t_S$ depend on the regions where $(t,x)$ lies in:
\beq
t_S =
\left\{
\barr{ll}
t_p ,& \mbox{if $(t,x)$ lies in region I or II},\\
t_q, & \mbox{if $(t, x)$ lies in region III or IV},
\earr
\right.
\eeq
and
\beq
x_S =
\left\{
\barr{ll}
x_0+(t - t_0)c+\sigma,& \mbox{if $(t, x)$ lies in region I or IV},\\
x_0+(t - t_0)c-\sigma, & \mbox{if $(t, x)$ lies in region II or III}.
\earr
\right.
\eeq

Since the transformed ($\tau,\xi$) domain is free space, we have the constitutive equations $D=\epsilon E$ and $B = \mu H$.
The corresponding constitutive equations in ($t,x$) domain are magneto-electric:
\bea
E_z &=& \alpha D_z + \beta B_y, \label{Econst}\\
H_y &=& \beta D_z + \gamma B_y, \label{Hconst}
\eea
where $\alpha$, $\beta$, and $\gamma$ are space and time dependent where
\beq
\left\{
\barr{rcl}
\alpha 	&=& 1/a_{12},						\\
\beta		&=& -a_{22}/a_{12} ,					\\
\gamma 	&=& (a_{12}a_{21}-a_{11}a_{22})/a_{12},	
\earr
\right.
\label{eq:cpar}
\eeq
and
\beq
\mtwo{cc}{a_{11}&a_{12}}{a_{21}&a_{22}}
	= \lam^{-1}\mtwo{cc}{0&\eps}{1/\mu&0} \lam.
\eeq

\section{The conventional FDTD method for time dependent magneto-electric medium}

Consider the following 1+1 D Maxwell's equations
\bea
\pd{D_z}t 	&=&{\pd{H_y}x}, \\
\pd{B_y}t 	&=&{\pd{E_z}x},
\eea
where the constitutive equations are given in Eqs. (\ref{Econst}) and (\ref{Hconst}).
For simplicity, we let $E = E_z$, $D = D_z$, $H = H_y$, and $B = B_y$.
We note that the primary issue with simulating magneto-electric media arises from the more complicated constitutive relations given in Eqs.
 (\ref{Econst}) and (\ref{Hconst}).
Using the conventional FDTD method, the discretized Maxwell's equations (the $D$ first scheme) are
\begin{equation}\label{YeeD}
D_{i}^{n+\frac{1}{2} } = D_{i}^{n-\frac{1}{2} } + \frac{\Delta t}{\Delta x}  \left  (H_{i+\frac{1}{2}}^{n} - H_{i-\frac{1}{2}}^{n} \right),
\end{equation}
\begin{equation}\label{YeeB}
B^{n+1}_{i+\frac{1}{2}} = B^{n}_{i+\frac{1}{2}} + \frac{\Delta t}{\Delta x} \left (E^{n+\frac{1}{2}}_{i+1} - E^{n+\frac{1}{2}}_{i} \right ),
\end{equation}
and the constitutive relations (assuming that the coupled constitutive relations respond locally in space and time) become
\begin{equation} \label{Econstdis}
E^{n+\frac{1}{2}}_{i} = \alpha^{n+\frac{1}{2}}_{i} D^{n+\frac{1}{2}}_{i} + \beta^{n+\frac{1}{2}}_{i} B^{n+\frac{1}{2}}_{i},
\end{equation}
\begin{equation} \label{Hconstdis}
H^{n+1}_{i+\frac{1}{2}} = \beta^{n+1}_{i+\frac{1}{2}} D^{n+1}_{i+\frac{1}{2}} + \gamma^{n+1}_{i+\frac{1}{2}} B^{n+1}_{i+\frac{1}{2}}.
\end{equation}

Parameters $\alpha$, $\beta$, and $\gamma$ are computed analytically from Eq.
(\ref{eq:cpar}) and are thus available at any point $(t,x)$ in the computational domain.
However, the quantities $B^{n+\frac{1}{2}}_{i}$ in Eq. (\ref{Econstdis}) and
$D^{n+1}_{i+\frac{1}{2}}$ in Eq. (\ref{Hconstdis}) are not computed in the discretized Maxwell's Eqs. (\ref{YeeD}) and (\ref{YeeB}).
%If a "B first" Yee scheme is used similar issues arise with the quantities $H^{n}$ and  $E_{i}$.
As a result, the conventional Yee approach cannot be directly used in the simulation of magneto-electric media.

One method for resolving these issues is to use time extrapolation and space interpolation.
To compute $B^{n+\frac{1}{2}}_{i}$ in Eq. (\ref{Econstdis}),  we have,
\bea
B^{n+\frac{1}{2}}_{i} &\approx& \frac{3B^{n}_{i}-B^{n-1}_{i}  }{2}\\
&\approx& \frac{\frac{3}{2}(B^n_{i+\half}+B^n_{i-\half}) - \half(B^{n-1}_{i+\half}+B^{n-1}_{i-\half})}{2} \\
&\approx& \frac{ 3B^{n}_{i+\frac{1}{2} }+ 3B^{n}_{i - \frac{1}{2}} - B^{n-1}_{i+\frac{1}{2}}-B^{n-1}_{i-\frac{1}{2}} } {4},
\eea
which is valid after the first timestep.  For the initial timestep, we use $B^{n}_{i}$ as an approximation for $B^{n+\frac{1}{2}}_{i}$.
Similarly, we compute $D^{n+1}_{i+ \frac{i}{2}}$ in Eq. (\ref{Hconstdis}) as follows

\begin{equation} \label{dapp}
D^{n+1}_{i+\half} \approx \frac{3D^{n+\frac{1}{2}}_{i+1}+3D^{n+\frac{1}{2}}_{i} - D^{n-\frac{1}{2}}_{i+1}-D^{n-\frac{1}{2}}_{i}}{4}.
\end{equation}

Numerical simulation on spacetime cloak shows that the conventional FDTD method presented in this section is unstable. This is primarily due to the use of time extrapolation. To avoid such extrapolation, we propose an overlapping Yee algorithm for time dependent magneto-electric media as described in the next section.

\section{The overlapping Yee Algorithm for time dependent magneto-electric media}

Rather than a single $D$ first scheme or $B$ first scheme, we use a combination of both.  While computationally more expensive, this produces $D$ and $B$ values at every half time step and spatial step within our computational domain.  As a result, we are able to perform the required magneto-electric Maxwell constitutive relation updates without extrapolation.

\begin{figure}[t]
 \centering
\subfigure[]{ \includegraphics[width=0.4\linewidth]{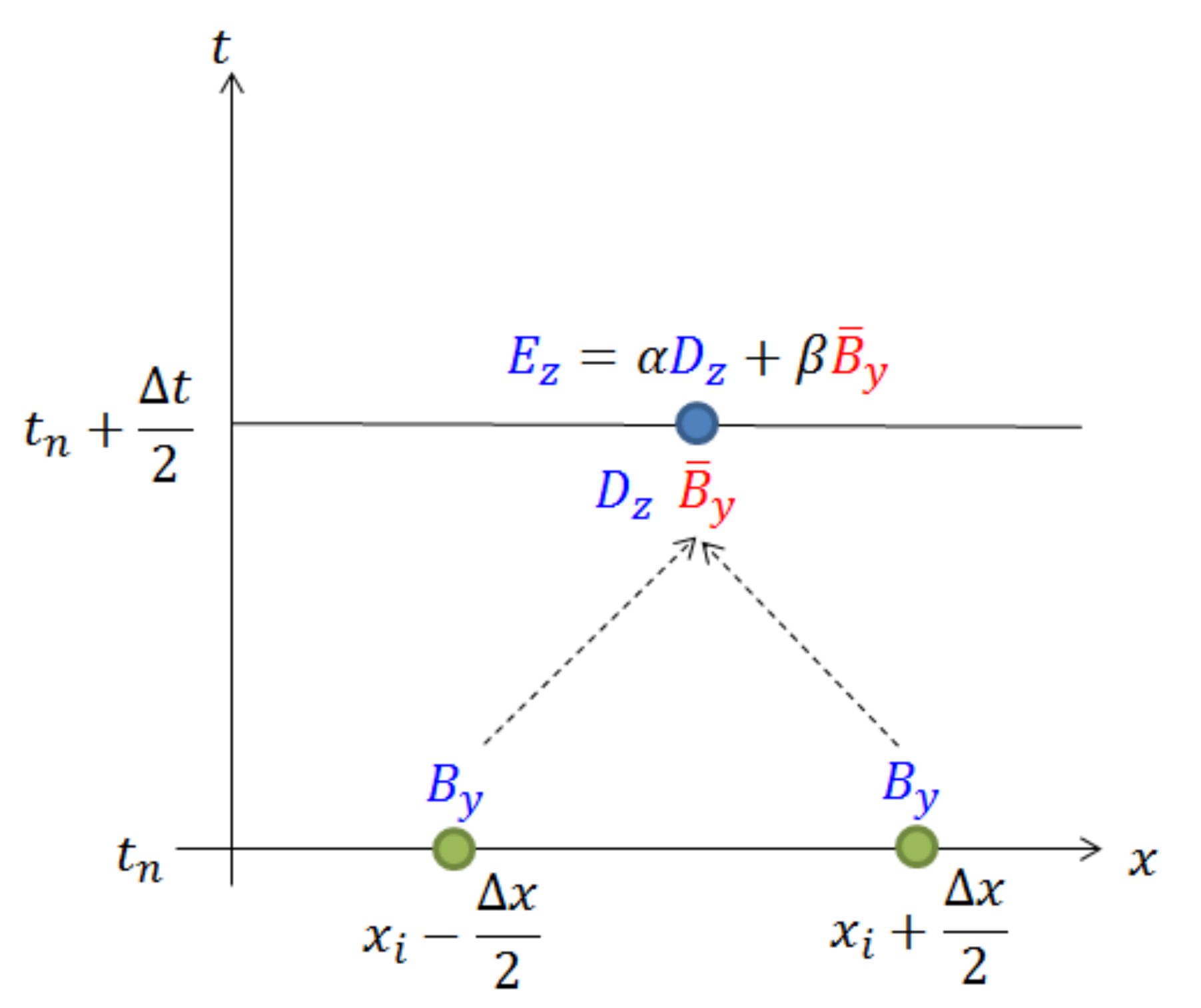} \label{fig:yee}}
\subfigure[]{ \includegraphics[width=0.4\linewidth]{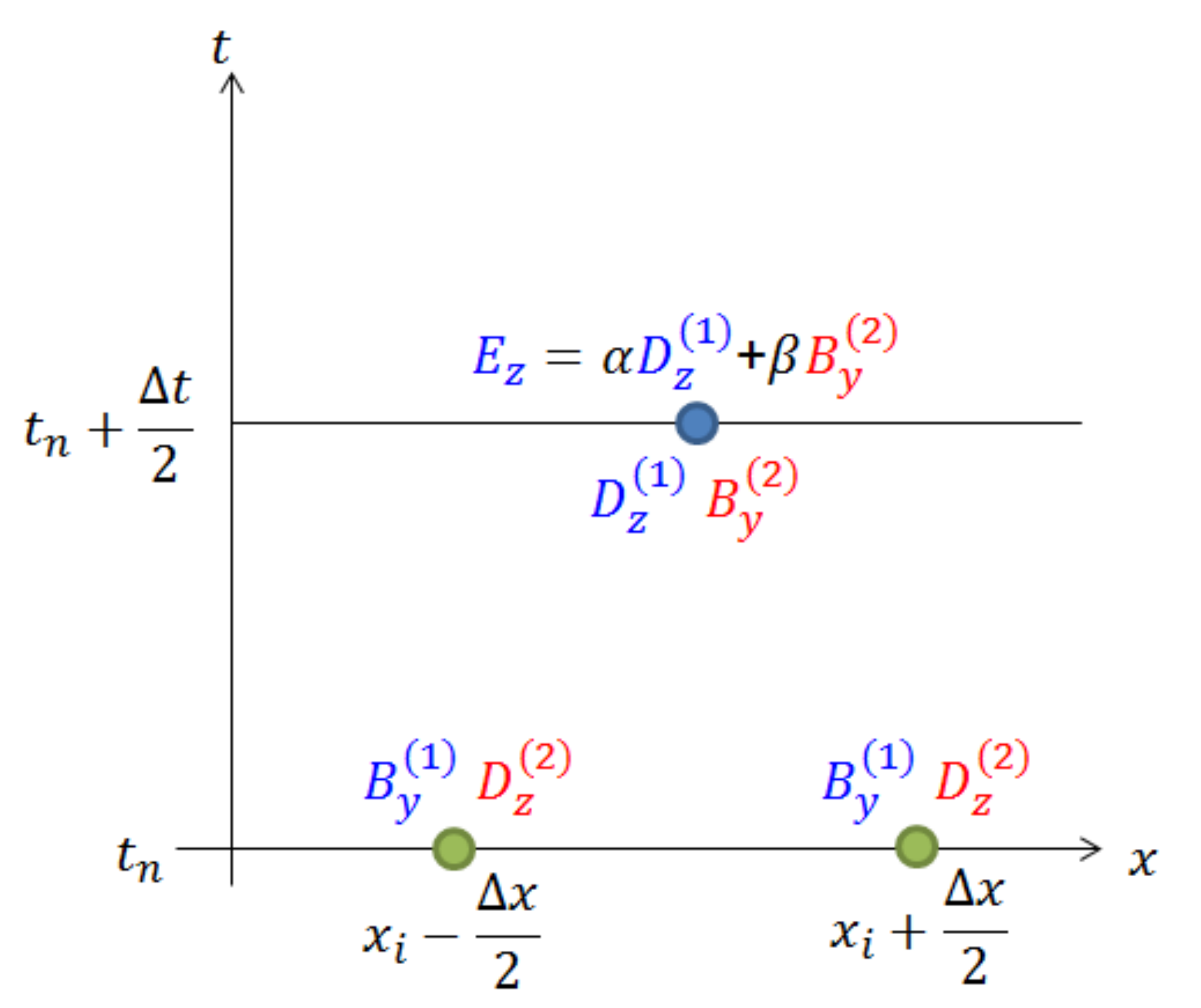} \label{fig:oy}}
 \caption{Computation of $E_z$ using (a) the time extrapolation under a conventional FDTD method and
(b) the overlapping Yee algorithm without extrapolation. }
\label{fig:grid}
\end{figure}

Figure~\ref{fig:grid} illustrates the fundamental differences between utilizing a conventional FDTD approach and the overlapping grid approach.
Figure~\ref{fig:yee} depicts that, due to the coupled constitutive relations, at any given step, $E_z$ has a dependence on the quantity $\overline{B_y}$.  The dotted arrows leading to this value indicate that $\overline{B_y}$ must be extrapolated in time and interpolated in space from values known on the conventional FDTD grid.

On the other hand, Fig.~\ref{fig:oy} illustrates that by utilizing a dual overlapping grid approach, the dependence on time extrapolation and spatial interpolation can be eliminated.   A new superscript notation is used to indicate the usage of two overlapping grids and to specify which grid each value of $B_y$ and $E_z$ comes from.  All values needed to compute $E_z$ at a particular instance in time exists on one of the two grids.

In presenting an algorithm for the adjusted scheme we utilize the convention that $n$ represents the time coordinate of the grid point and $i$ represents the spatial coordinate.  Indices of $n\pm \frac{1}{2}$ and $i \pm \frac{1}{2}$ are use to represent locations that are a half time step or spatial step from the gridpoint.  With this established, we present an algorithm for an overlapping Yee FDTD method.

The algorithm in 1D case can be summarized by the following steps:
\begin{enumerate}
\item {\it Update $D^{n+\frac{1}{2}}_{i}$ on the first Yee grid and $B^{n+\frac{1}{2}}_{i}$ on the second Yee grid:}
$$D^{n+\frac{1}{2}}_{i} = D^{n-\frac{1}{2}}_{i} + \frac{\Delta t}{\Delta x} \left[ H^{n}_{i+\frac{1}{2}} -  H^{n}_{i-\frac{1}{2}} \right],$$
$$B^{n+\frac{1}{2}}_{i} = B^{n-\frac{1}{2}}_{i} + \frac{\Delta t}{\Delta x} \left[ E^{n}_{i+\frac{1}{2}} -  E^{n}_{i-\frac{1}{2}} \right].$$
\item {\it Update $E^{n+\frac{1}{2}}_{i}$ and $H^{n+\frac{1}{2}}_{i}$:}
$$E^{n+\frac{1}{2}}_{i} = \alpha^{n+\frac{1}{2}}_{i} D^{n+\frac{1}{2}}_{i} + \beta^{n+\frac{1}{2}}_{i} B^{n+\frac{1}{2}}_{i},$$
$$H^{n+\frac{1}{2}}_{i} = \beta^{n+\frac{1}{2}}_{i} D^{n+\frac{1}{2}}_{i} + \gamma^{n+\frac{1}{2}}_{i} B^{n+\frac{1}{2}}_{i}.$$
\item {\it Update $B^{n+1}_{i+\frac{1}{2}}$ on the first Yee grid and $D^{n+1}_{i + \frac{1}{2}}$ on the second Yee grid:}
$$B^{n+1}_{i+\frac{1}{2}} = B^{n}_{i+\frac{1}{2} } + \frac{\Delta t}{\Delta x} \left[ E^{n+\frac{1}{2}}_{i+1} - E^{n+\frac{1}{2}}_{i} \right ],$$
$$D^{n+1}_{i + \frac{1}{2}} = D^{n}_{i+\frac{1}{2} } + \frac{\Delta t}{\Delta x} \left[ H^{n+\frac{1}{2}}_{i+1} - H^{n+\frac{1}{2}}_i  \right ].$$
\item {\it Update $E^{n+1}_{i+\frac{1}{2}}$ and $H^{n+1}_{i+\frac{1}{2}}$:}
$$E^{n+1}_{i+\frac{1}{2}}  = \alpha^{n+1}_{i+\frac{1}{2}} D^{n+1}_{i+\frac{1}{2}}   +\beta^{n+1}_{i+\frac{1}{2}} B^{n+1}_{i+\frac{1}{2}},$$
$$H^{n+1}_{i+\frac{1}{2}}  = \beta^{n+1}_{i+\frac{1}{2}} D^{n+1}_{i+\frac{1}{2}}   +\gamma^{n+1}_{i+\frac{1}{2}} B^{n+1}_{i+\frac{1}{2}}.$$
\end{enumerate}

We note that the extension to 2-D and 3-D should be feasible by utilizing a time collocated grid (where $B$ and $E$ are computed at the same instance in time, but not necessarily in space).  Doing so eliminates time extrapolation in 2-D and 3-D but still uses interpolation in space.

\section{Spacetime cloak simulation}

The simulation of magneto-electric media has direct applications to the spacetime cloak presented in \cite{McCall11}.  We create a similar diamond shaped spacetime cloak using the transformation depicted in Fig.~\ref{fig:tran}.
In simulating the above cloak, we utilize a 10~$\mu m$ by 15~$fs$ long computational domain with a plane wave (with $\lambda =  600$~$nm$) traveling in the positive $x$ direction.
Our spacetime cloak is 1200~$nm$ wide and lasts 6~$fs$.  With regards to Fig. \ref{fig:tranxt} we have $x_0 = 1.5$~$\mu m$, $t_0 = 9$~$fs$, $t_q = 6$~$fs$, $t_p=12$~$fs$, $\sigma = 600$~$nm$ and $\delta = 180$~$nm$.
The computational domain is terminated by perfectly matched layer (PML) absorbing boundary conditions \cite{Taflove05}.
Due to the fast and slow phase velocities in the spacetime cloak \cite{McCall11}, $\Delta t$ is chosen by
$\Delta t \le \Delta x / v_{max}$ where $v_{max}$ is the largest wave speed computed in the cloaking region.
We perform the simulation 3 times with the number of spatial cells, $N$, equal to 1600, 2000, and 2400 and the same number of time steps

\begin{figure}
 \centering
\subfigure[]{ \includegraphics[width=0.40\linewidth]{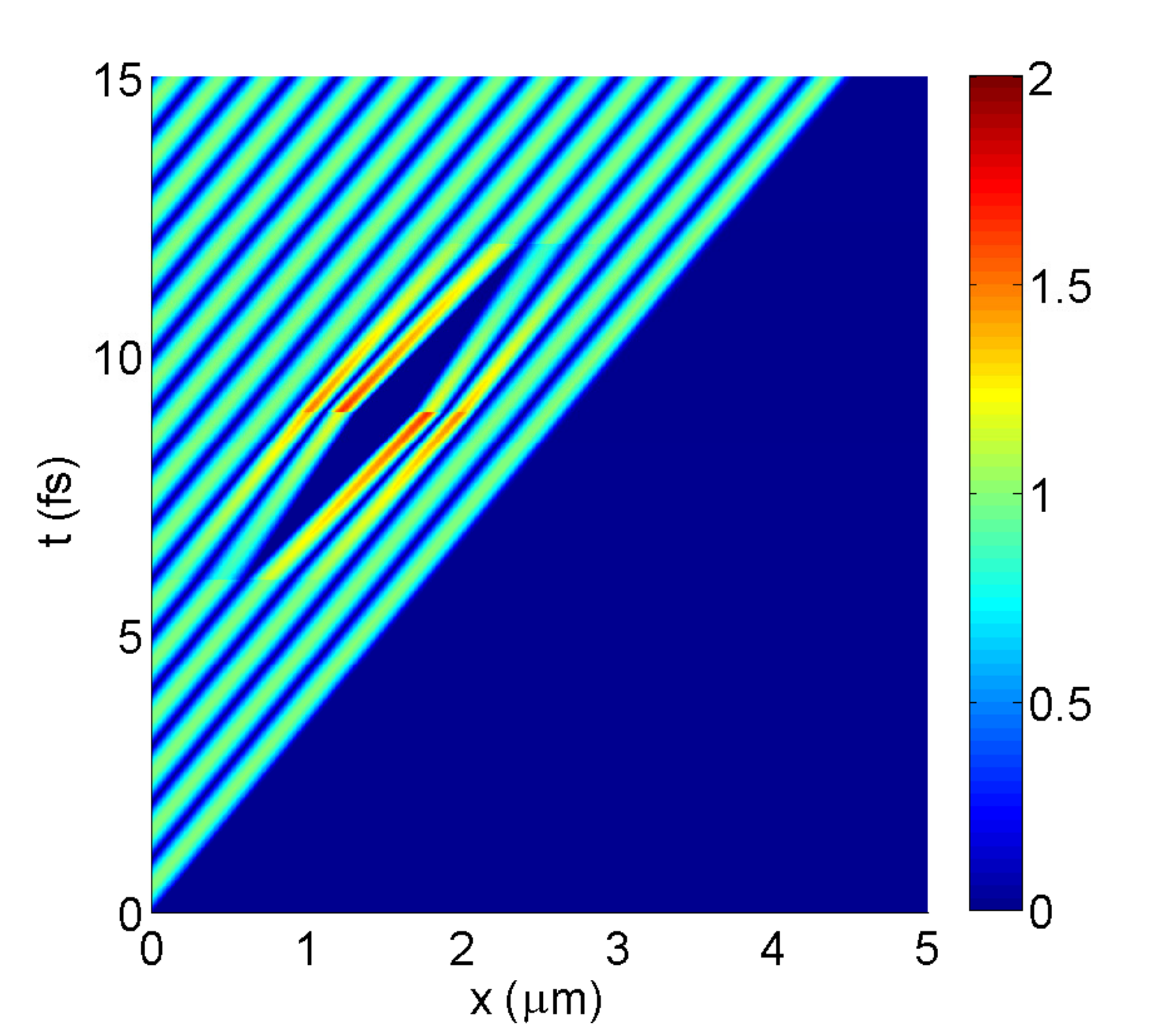}  }
\subfigure[]{ \includegraphics[width=0.40\linewidth]{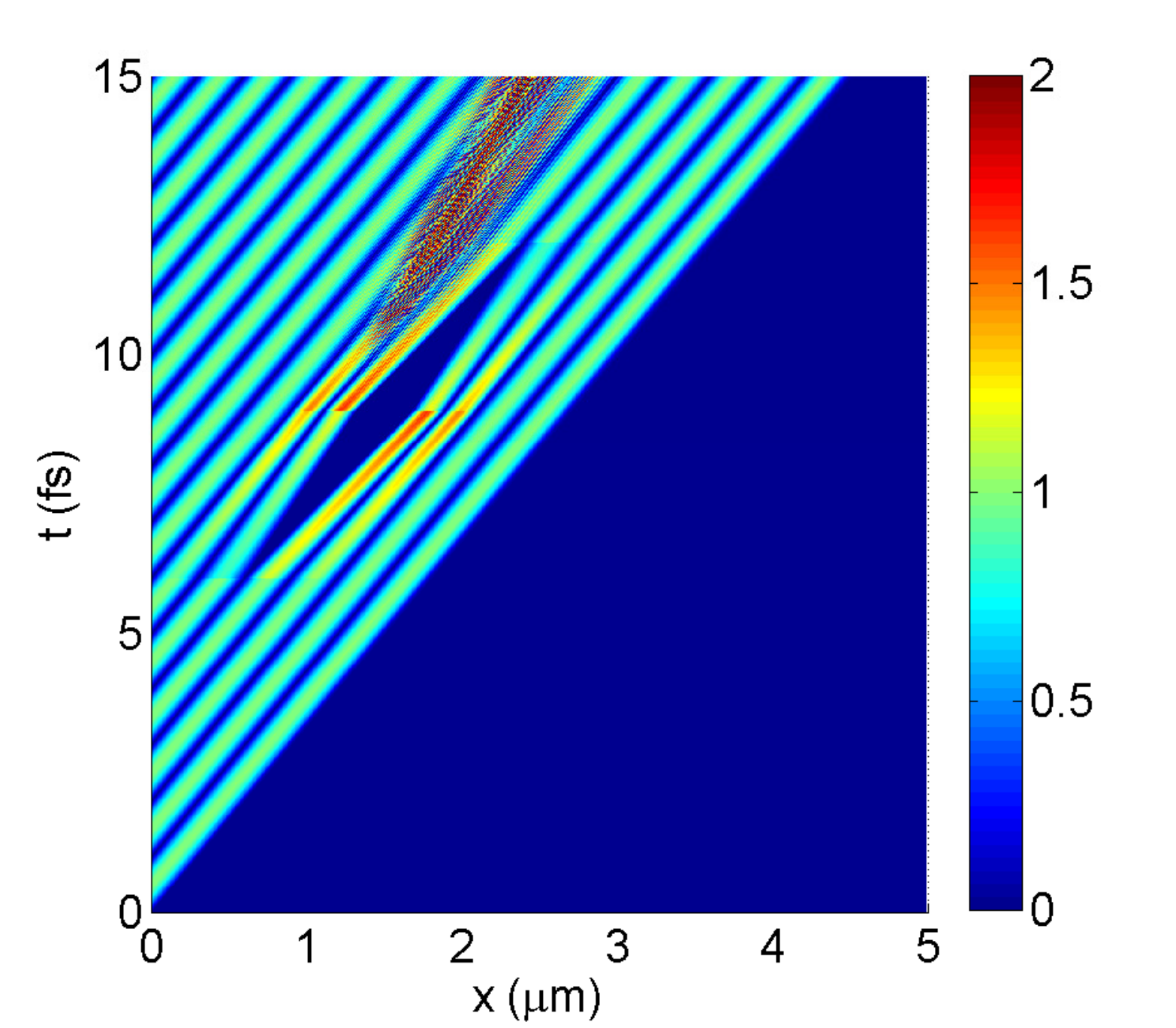} } \\
\subfigure[]{ \includegraphics[width=0.40\linewidth]{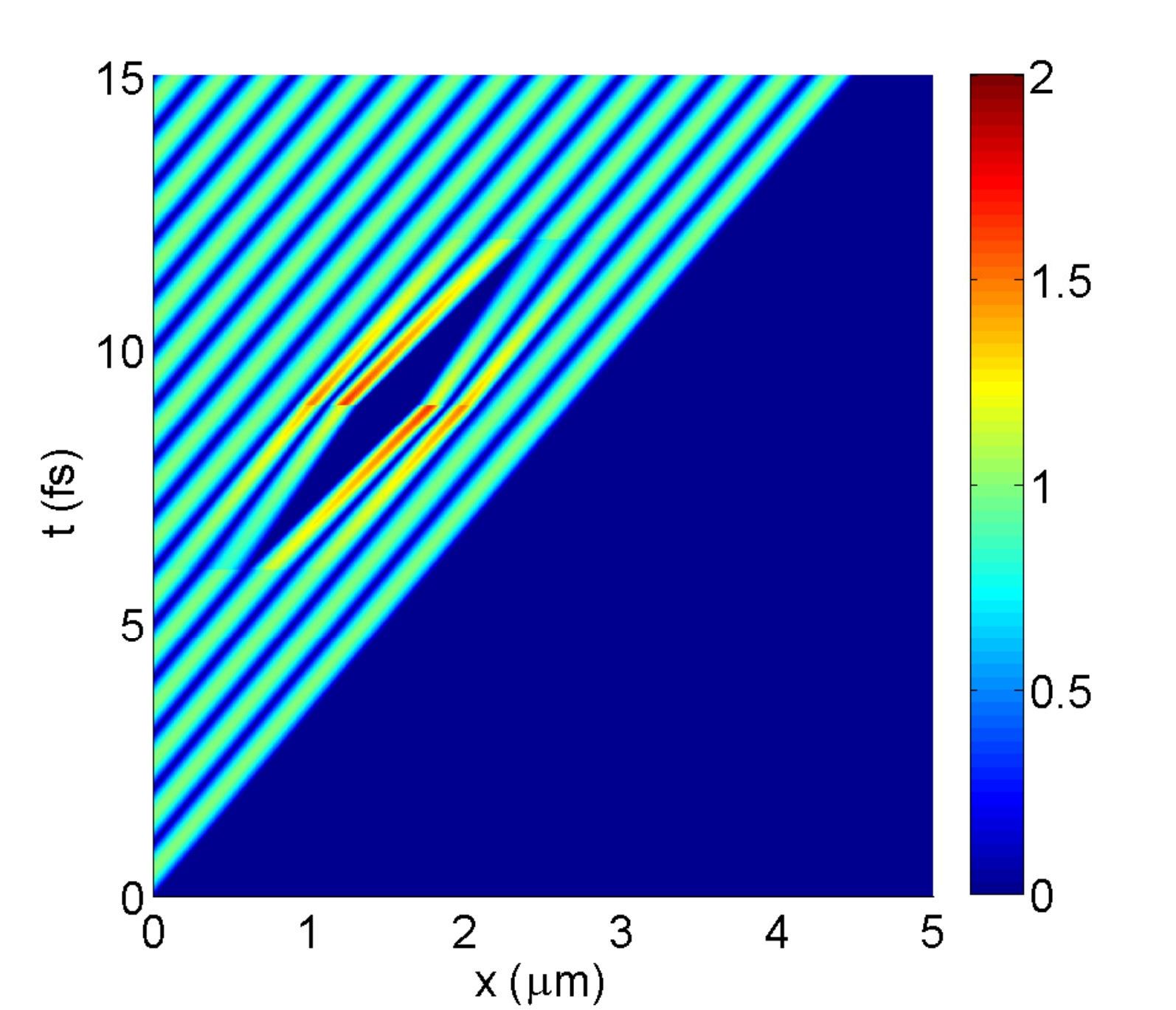}  }
\subfigure[]{ \includegraphics[width=0.40\linewidth]{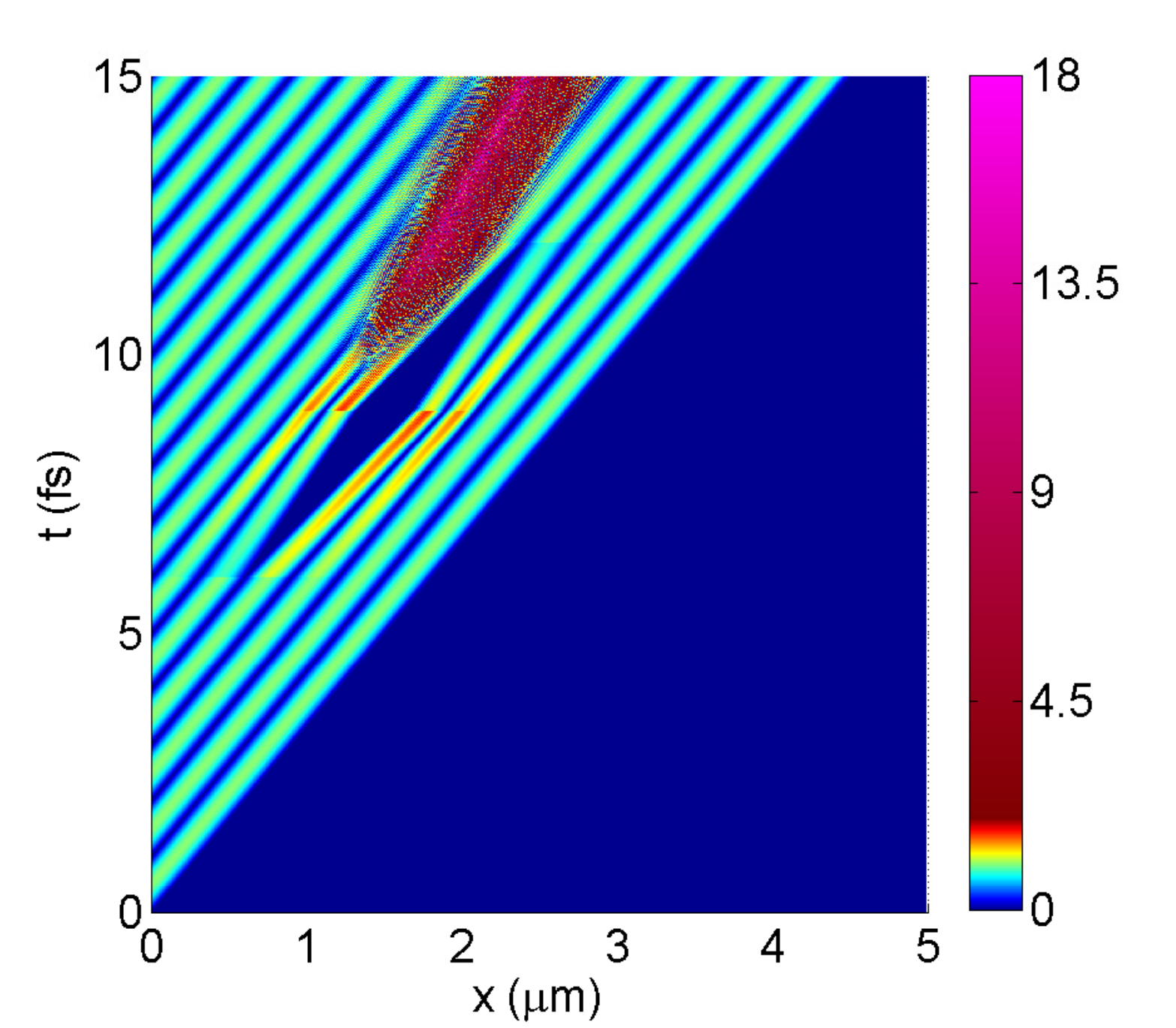}}\\
\subfigure[]{ \includegraphics[width=0.40\linewidth]{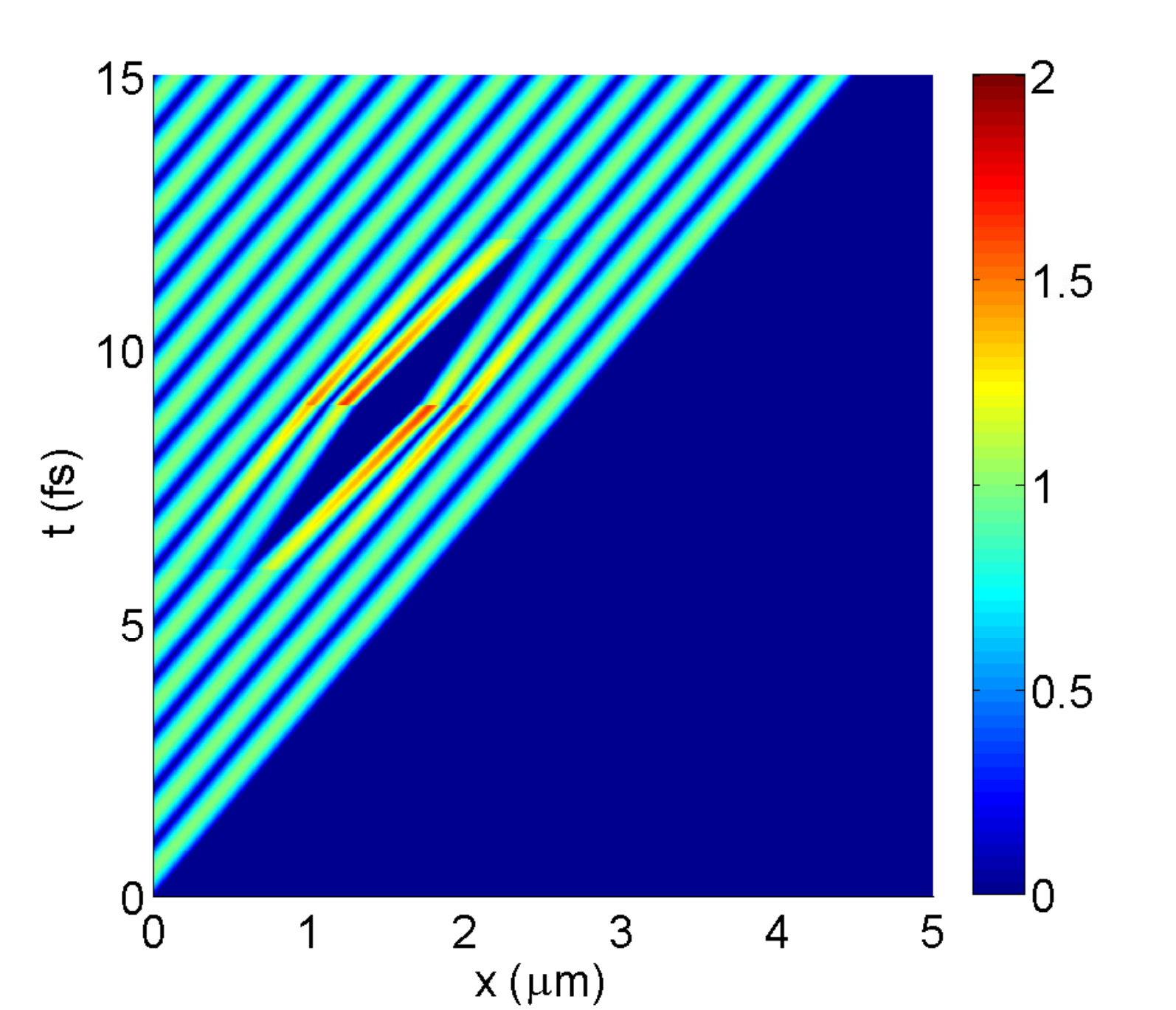}  }
\subfigure[]{ \includegraphics[width=0.40\linewidth]{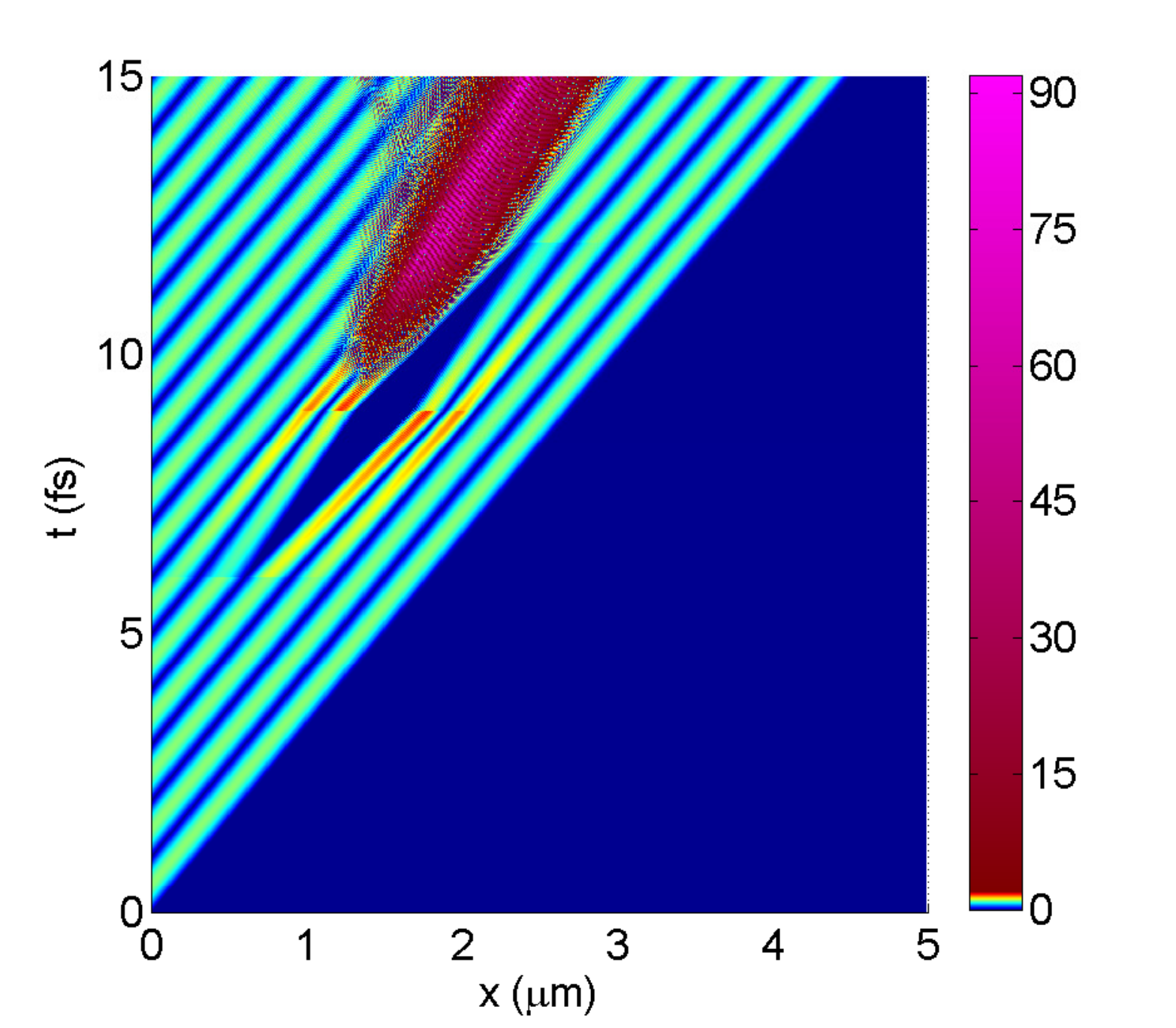} }
 \caption{Contour plots of the electric field intensity for spacetime cloak simulations.
Left column: the overlapping Yee FDTD results with grid sizes (a) $N = 1600$, (c) $N = 2000$, and (e) $N = 2400$, respectively.
Right column: the corresponding results using a conventional time extrapolation based FDTD method.}
\label{fig:stsim}
\end{figure}

The numerical simulation results are shown in Fig.~\ref{fig:stsim}.  The figures shown are contour plots of the intensity of the electric field generated by the interaction between the incident source and the spacetime cloak. We compare the conventional FDTD method with time extrapolation and the proposed overlapping Yee FDTD method for various spatial resolutions.  The result shows the bending of the electric field (and consequently the electromagnetic wave) in space and time around the cloaked event.
The shown simulation results verify that an extrapolation approach is unstable (near the closing process of the cloaked event) and the proposed overlapping FDTD approach provides stable results.

\begin{figure}[t]
\centering
\subfigure[Wave enters cloaking region]{ \includegraphics[width=0.40\linewidth]{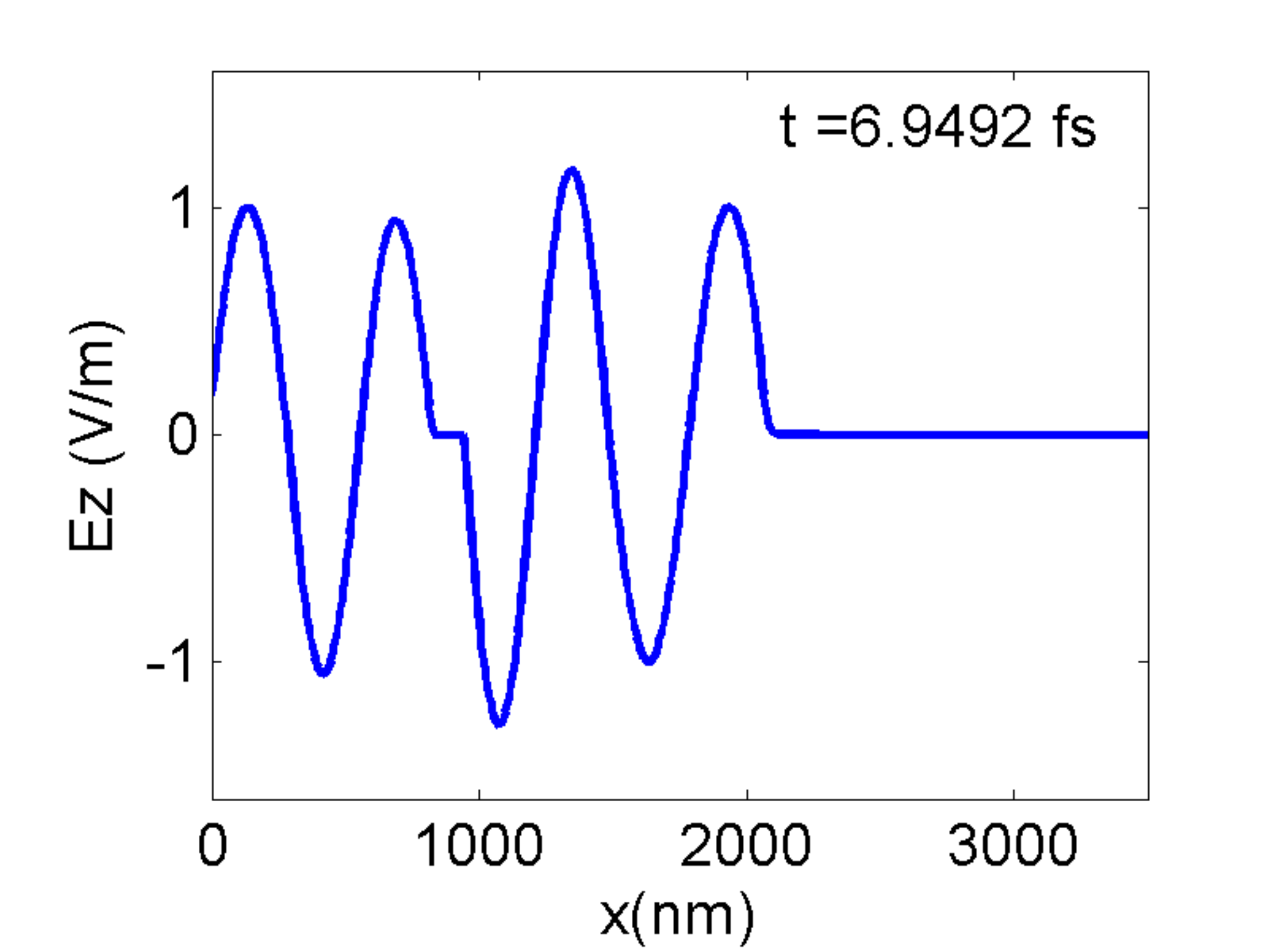} \label{fig:cloakopen}  }
\subfigure[Wave at center of cloak]{ \includegraphics[width=0.40\linewidth]{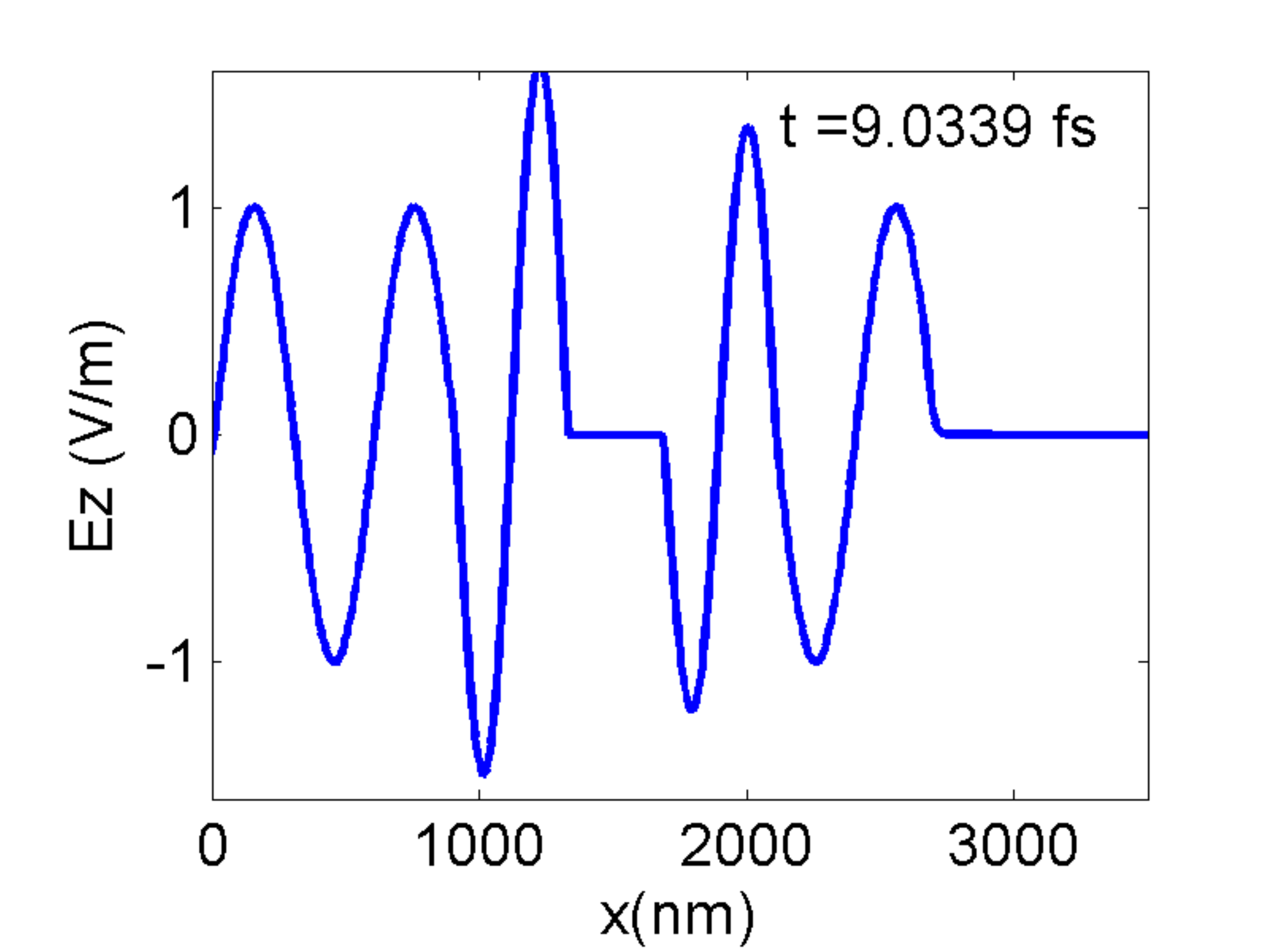} \label{fig:cloakcenter} } \\
\subfigure[Wave leaving cloaking region]{ \includegraphics[width=0.40\linewidth]{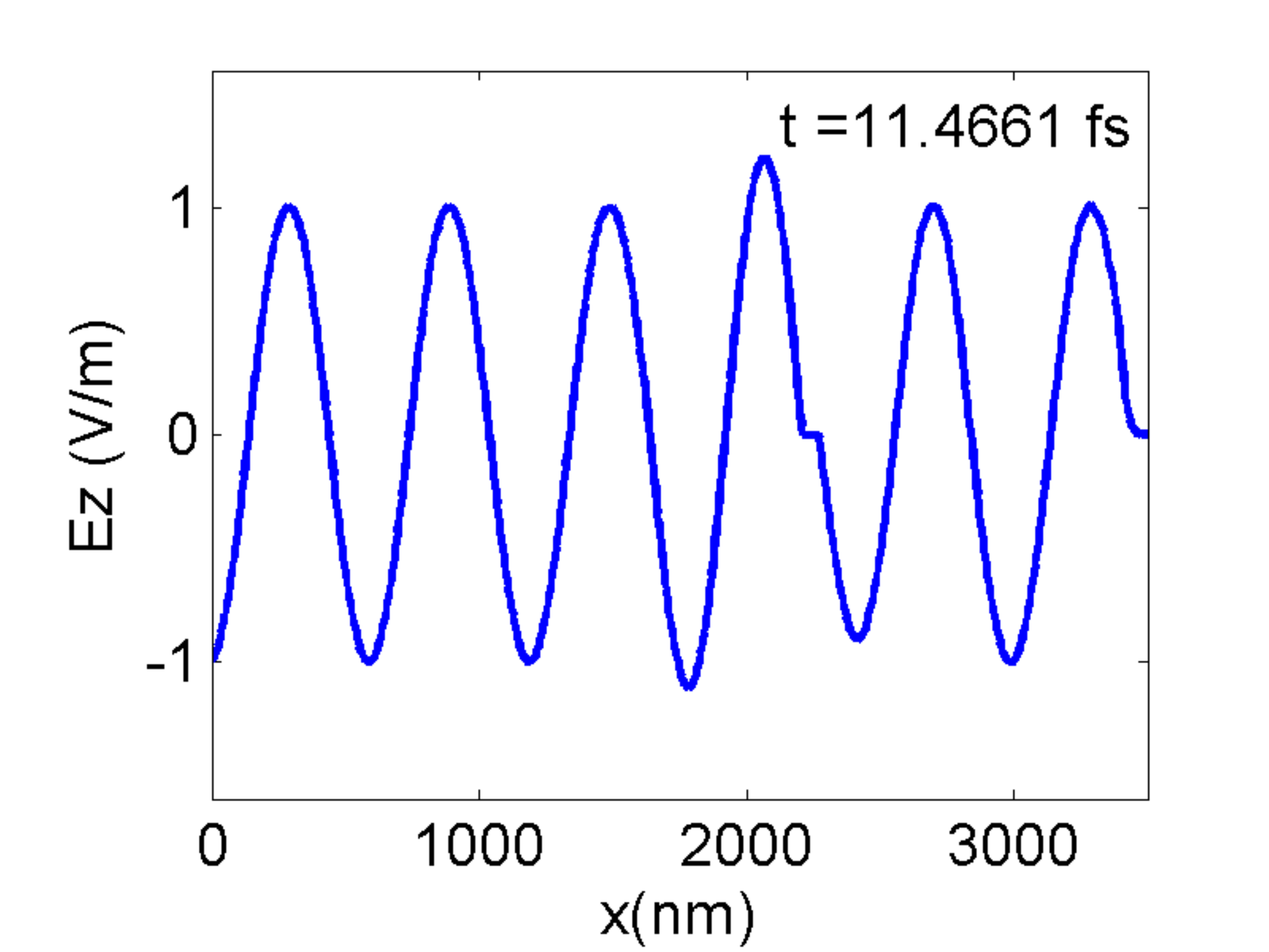}  \label{fig:cloakclose} }
\subfigure[Wave after leaving cloak]{ \includegraphics[width=0.40\linewidth]{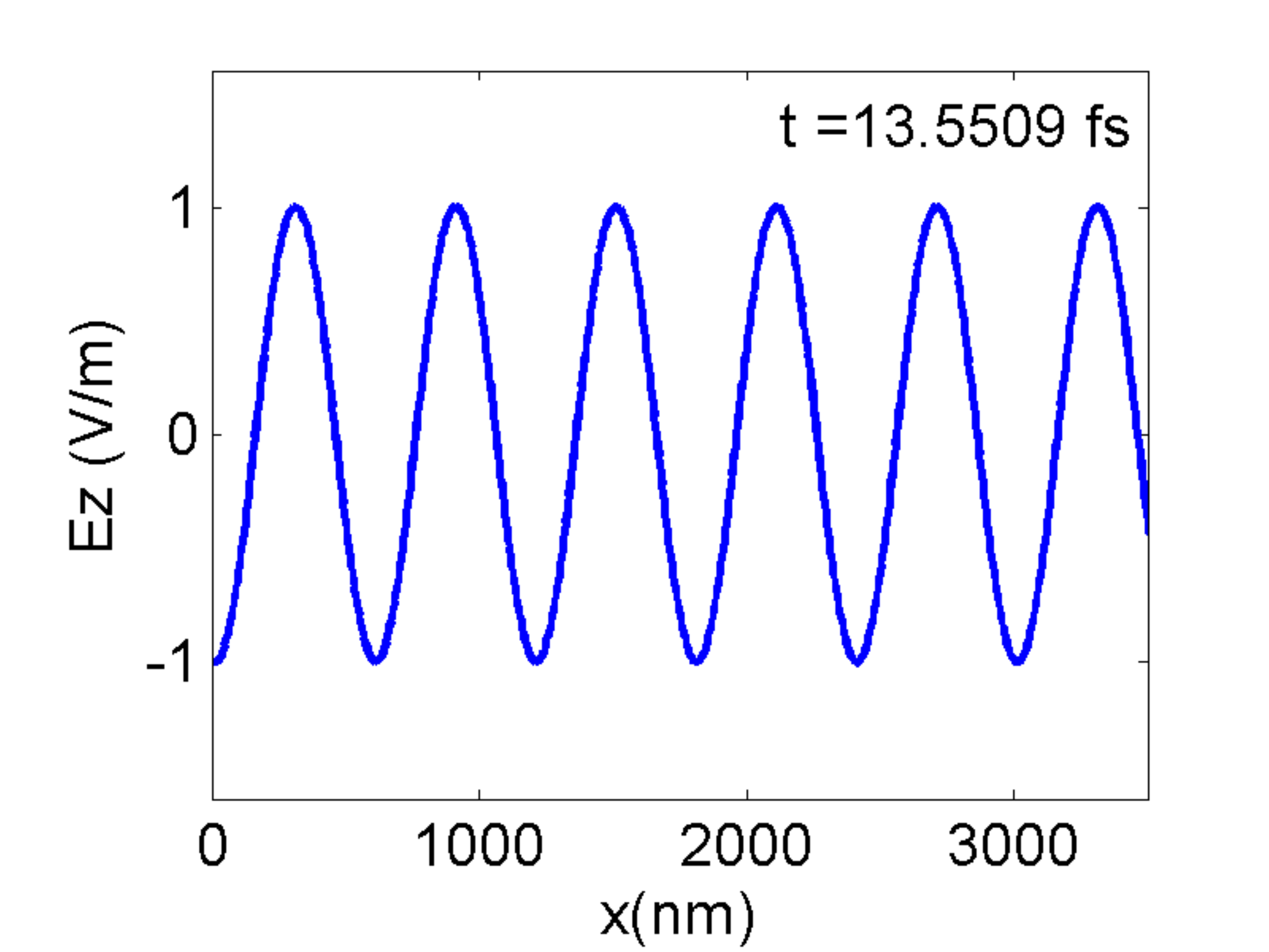} \label{fig:cloaklate} } \\
\caption{Evolution of the Electric field as the wave propagates through the spacetime cloak.}
\end{figure}

We now examine the behavior of the electric field as the wave propagates through the diamond shaped space-time region under the same simulation parameters.  As the wave enters the cloak it begins to split as the front portion of the wave speeds up and the back portion of the wave slows down.  This effect is made possible due to the space-time dependence of the media within the cloaking region. As this separation occurs, it is mirrored in the electric field producing a gap in the electric field.  The early development of this gap is shown in Fig.~\ref{fig:cloakopen} near $x = 1000~nm$.  As the wave progresses through the media, the cloak widens in the $x$ direction allowing for the size of this gap to increase as time progresses.  The gap in the electric field, as shown in Fig.~\ref{fig:cloakcenter}, is at a maximum at the center of the cloaking region. As the wave propagates past this point, the cloak (shrink in the $x$ direction) begins to close.  Due to the closing of the cloak, the front of the wave begins to slow down while the back speeds up.  This action allows for the wave to remerge just as it exits the cloak.  The field behavior moments before the wave exits the cloak is shown in Fig.~\ref{fig:cloakclose}. At this point a small gap near $x = 2200~nm$ still exists.  Once the wave has propagated through the cloak region, it is fully rejoined and the propagation behavior and resulting field behavior returns to that of propagation in an isotropic media, as shown in Fig.~\ref{fig:cloaklate}.

\section{Stability Analysis}

\begin{figure}[t]
\centering
\subfigure[]{ \includegraphics[width=0.45\linewidth]{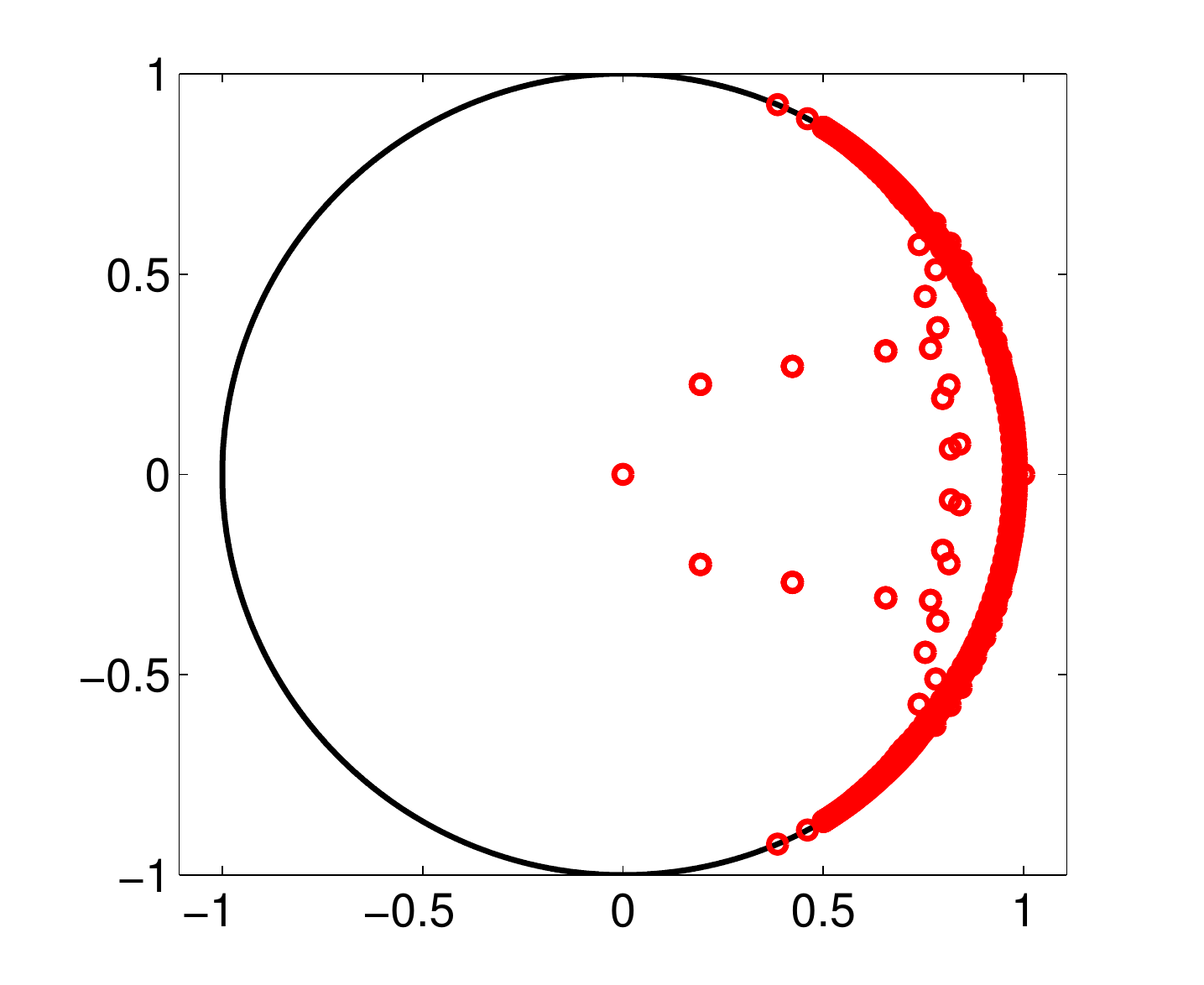}  \label{fig:overlapeigs}}
\subfigure[]{ \includegraphics[width=0.45\linewidth]{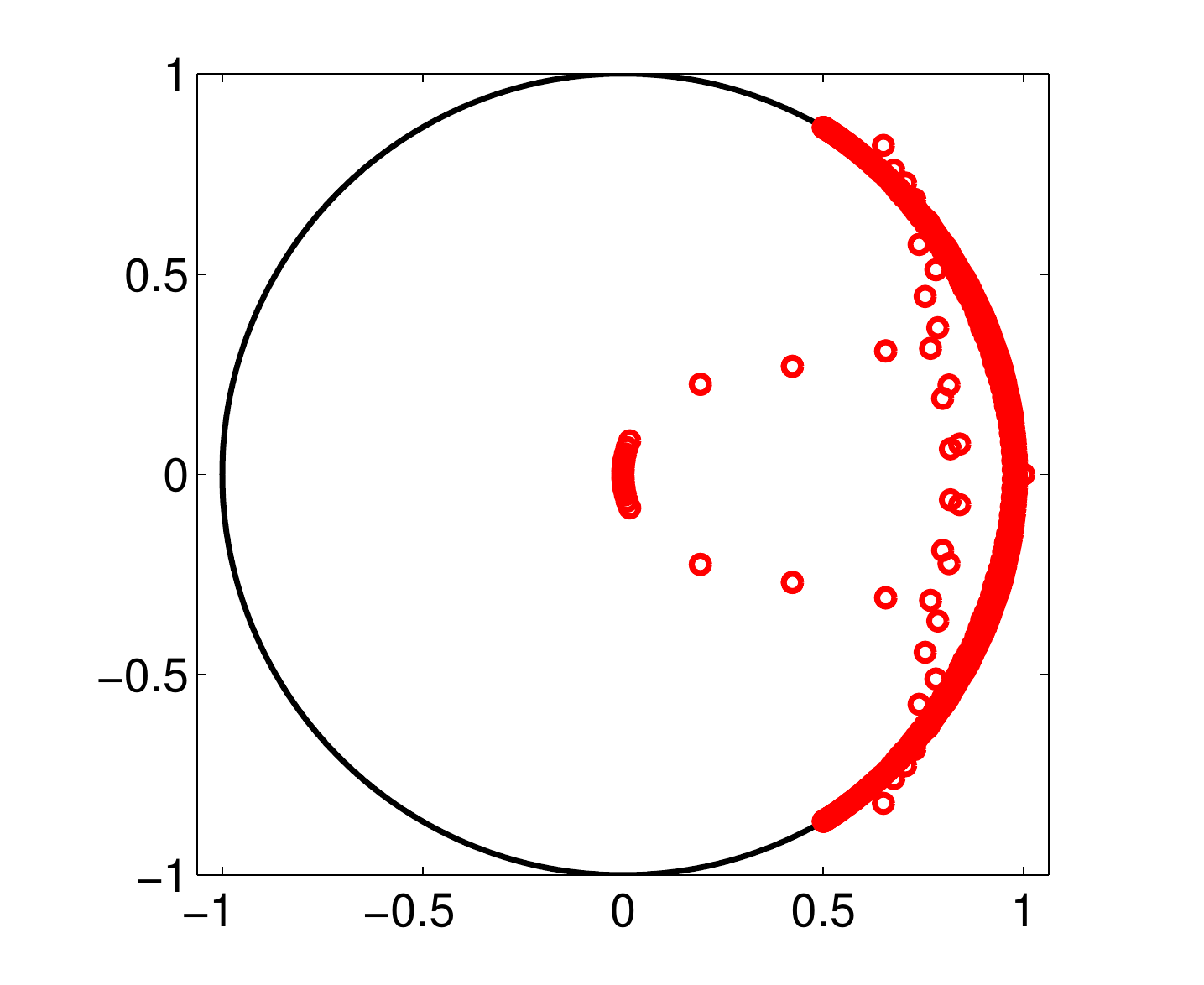} \label{fig:yeigs}} \\
\caption{Distribution of the eigenvalues in the complex plane for (a) the overlapping Yee FDTD method and (b) the conventional FDTD method.}
\label{fig:eigs}
\end{figure}

To conduct a stability analysis, we compare the distribution of eigenvalues of the conventional FDTD method to that of our overlapping grid approach with respect to the unit circle.  For both tests we utilize similar simulation parameters as in the previous section, but with a smaller mesh size of $N = 200$. We compute the eigenvalues of the numerical update equations at a time instant when $t \in (t_q, t_p)$ and plot them in the complex plane. The results when $t = 8.34~fs$ are shown in Fig.~\ref{fig:eigs}.
%Similar results are obtained for other time instants when $t \in (t_q, t_p)$.

As indicated in the Fig.~\ref{fig:overlapeigs} all the eigenvalues of the overlapping approach lie within or on the unit circle with the ones lying on unit circle being simple eigenvalues.  On the other hand, Fig.~\ref{fig:yeigs} shows that some of eigenvalues (near the upper and lower right corners) of the conventional FDTD approach lie outside the unit circle.  In particular the modulus of the largest eigenvalue is 1 for the overlapping method and 1.046 for the conventional FDTD method.  This indicates that the overlapping Yee FDTD method is stable while the conventional FDTD approach (with extrapolation) would experience instability.
%These stability results agree with the results shown in Fig.~\ref{fig:stsim}.

\section{Conclusion}
In conclusion, we have proposed a stable FDTD method for simulation of space and time dependent magneto-electric medium based on the use of two sets of overlapping Yee grids that are offset in time by a half time step.
Additionally, we have shown the direct application of this method to the simulation of the spacetime cloak.
The proposed method is useful in exploration of other new physical possibilities offered by metamaterials and transformation optics.

\section*{Acknowledgments}
We thank Dr. Yong Zeng for invaluable discussions.
This work was supported in part by the US Air Force Office of Scientific Research Grant FA9550-10-1-0127, US ARO Grant W911NF-11-2-0046, and NSF Grant HRD-1242067. M. Brio was also supported by US AFOSR MURI Grant FA9550-10-1-0561.

\end{document}